# Did your child get disturbed by an inappropriate advertisement on YouTube?

Jeffrey Liu[1], Rajat Tandon[1], Uma Durairaj[1], Jiani Guo[1], Spencer Zahabizadeh[1], Sanjana Ilango[1],
Jeremy Tang[1], Neelesh Gupta[1], Zoe Zhou[1], Jelena Mirkovic[2]

[1]University of Southern California, Los Angeles, California, USA
[2]USC Information Sciences Institute, Marina del Rey, California, USA
{jliu5021,rajattan,uduraira,jennyguo,szahabiz,ilango,tangjere,neeleshg,zoez}@usc.edu,mirkovic@isi.edu

**Abstract**

YouTube is a popular video platform for sharing creative content and ideas, targeting different demographics. Adults, older children, and young children are all avid viewers of YouTube videos. Meanwhile, countless young-kid-oriented channels have produced numerous instructional and age-appropriate videos for young children. However, inappropriate content for young children, such as violent or sexually suggestive content, still exists. And children lack the ability to decide whether a video is appropriate for them or not, which then causes a huge risk to children's mental health. Prior works have focused on identifying YouTube videos that are inappropriate for children. However, these works ignore that not only the actual video content influences children, but also the advertisements that are shown with those videos.

In this paper, we quantify the influence of inappropriate advertisements on YouTube videos that are appropriate for young children to watch. We analyze the advertising patterns of 24.6 K diverse YouTube videos appropriate for young children. We find that 9.9% of the 4.6 K unique advertisements shown on these 24.6 K videos contain inappropriate content for young children. Moreover, we observe that 26.9% of all the 24.6 K appropriate videos include at least one ad that is inappropriate for young children. Additionally, we publicly release our datasets and provide recommendations about how to address this issue.

## 1 Introduction

YouTube is one of the most widely used video sharing platforms, and it appeals to a large and diverse audience. There are about 2.1 billion users on YouTube, and over a billion hours of content have been uploaded worldwide [10]. Young children under the age of 6 are a significant portion of the YouTube audience. According to a study by Pew Research [26], out of 4,591 participants who are parents of young children, 81 percent of them agreed that their children watch videos on YouTube. Moreover, as per a new report from Common Sense Media [9], the amount of time that young children spend watching online videos has doubled since 2017.

Thus, with the tremendously increasing size of the young audience that consumes video content, numerous young-kid-oriented YouTube channels are trending. For example, Blippi [3], an educational channel for kids, has over 15 M subscribers. Admittedly, young-kid-oriented channels have a plethora of educational and meaningful videos. For example, the Blippi channel teaches kids numbers, colors and handcraft. But videos on such young-kid-targeted channels may be displayed along with inappropriate advertisements (ads for short) that present sexual, violent, drug and alcohol content [29]. And these inappropriate ads have raised serious concerns about young children's mental and physical health [20]. In 2018, a mother alerted the founder of a kids safety website *pedimom.com* about a disturbing ad that she accidentally watched [31] on a young-kid-oriented YouTube video. The ad was about an adult male walking on screen and teaching kids how to cut their wrists. Such horrifying ads put young children at risk. In addition, young children lack the critical thinking ability to determine if a certain piece of content is appropriate or not [8]. Although, in 2015, YouTube launched another platform, YouTube Kids, targeting ad-free content for the kids' audiences, studies show that kids still watch regular YouTube more than YouTube Kids [7, 11]. Meanwhile, YouTube Kids was not as safe an alternative as it claimed to be. As of May 2022, YouTube Kids still shows numerous videos promoting drug culture and firearms to young kids [17]. Similarly, YouTube's inappropriate ad filters for children require significant improvements. In 2020, Common Sense Media [9] surveyed 191 parents and found that out of the 1.6 K videos their children watched, 20% of them included inappropriate ads for children. Most of the ads were about physical violence. Our findings complement their observations about inappropriate ads being shown to young kids. However, there are noticeable differences as their study focuses on a relatively small-sized dataset, while our study is based on a more recent and 15 times larger dataset.

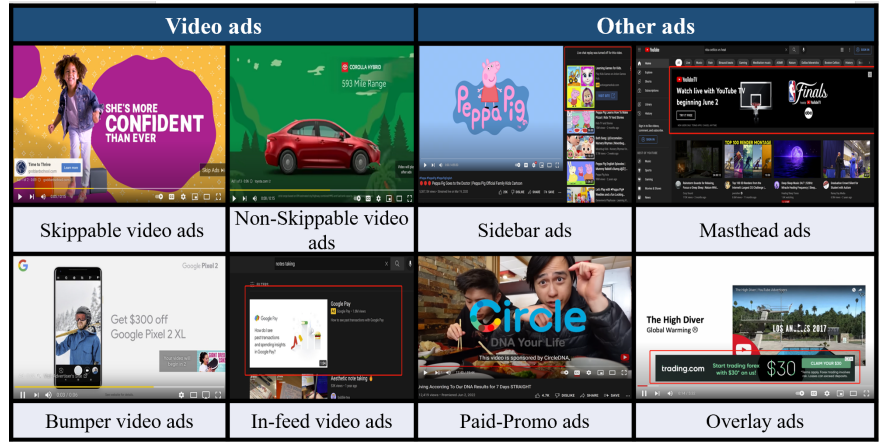

**Figure 1.** Examples of all types of YouTube advertisements

In this paper, we present a large-scale study that demonstrates the seriousness of the issue regarding inappropriate ads shown on young-kid-oriented videos. We define a young kid as a child under the age of 6. Our contributions are summarized as follows:

1. We conduct an extensive study of ad patterns in young-kid-oriented videos to quantify the presence of inappropriate ads that are shown on appropriate videos on YouTube. We analyze the ads for 24.6 K YouTube videos that are safe for young children to watch. **We find that 9.9% of the 4.6 K unique advertisements shown on these 24.6 K videos contain inappropriate content for young children. Moreover, we observe that 26.9% of all the 24.6 K appropriate videos include at least one inappropriate ad for young children.**

2. We publicly release a comprehensive dataset of YouTube ads displayed in young-kid-oriented videos [22].

3. We offer recommendations for YouTube to address the issue of inappropriate ads shown to young children.

## 2 Related Work

Prior works focus on detecting inappropriate video content for kids on YouTube. Eickhoff and De Vries [12] utilize information from YouTube videos, such as the number of views and likes of a video, as features to build a binary classifier for suitable videos for children. Kaushal et al. [19] demonstrate a machine learning classifier that utilizes video, user, and comment-level features to identify users that intentionally promote disturbing videos. Papadamou et al. [24] build a classifier using video metadata to identify inappropriate content that targets children on YouTube. Tahir et al. [30] use audio and visual elements, video frames, embedded audio, and character motions as features to build a deep learning based classifier to identify inappropriate videos for kids. By investigating the profiles and comments of viewers on popular children-oriented channels on YouTube, Araújo et al. [6] conclude that children under the age of 13 are easily exposed to inappropriate information and ads. Singh et al. [25] utilize video frames as the input to build a recurrent deep network-based classifier for detecting unsafe videos for kids. Ishikawa et al. [18] develop a static and motion-based deep learning classifier for identifying disturbing animated videos. In 2020, Common Sense Media [9] collected 1,639 YouTube videos viewed by 0 to 8-year-olds and found that one-fifth (326) of these videos contained age-inappropriate ads.

Our work is different from the above works, as our paper quantifies inappropriate ads shown on YouTube videos that are suitable for young kids, instead of classifying the videos themselves. Moreover, our work is the first quantitative study that analyzes inappropriate ad patterns from a large dataset (24.6 K) of young-kid-oriented videos.

## 3 Methodology

This section discusses the methodology we employ to (1) collect and annotate a comprehensive dataset, D1, of young-kid-oriented YouTube videos, (2) crawl and collect the dataset, D2 and D3, that comprises the video ads and sidebar ads shown on D1, and (3) annotate datasets, D2 and D3, to identify ads that are (a.) appropriate for young children to watch, (b.) inappropriate for young children to watch, and (c.) irrelevant for young children to watch. Further, we analyze those ads that are labeled inappropriate or irrelevant to better characterize their content (See Figure 4 for the architecture of our methodology).

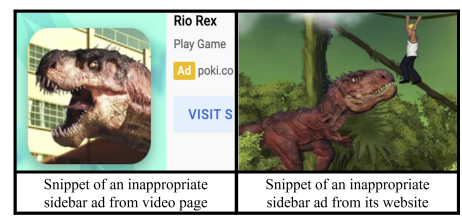

**Figure 2.** Snippet of an inappropriate sidebar ad

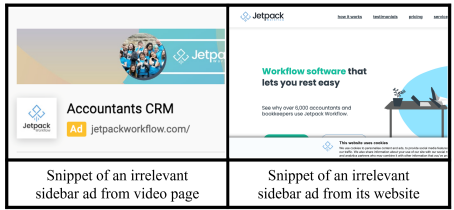

**Figure 3.** Snippet of an irrelevant sidebar ad

### 3.1 Definitions

We consult YouTube's guidelines [34] and the Children's Online Privacy Protection Act (COPPA) [13] to determine whether certain content is appropriate, inappropriate, or irrelevant for young audiences.

**Appropriate Content:** It includes suitable content for young kids and preschoolers (under the age of 6) and content that is relevant to their specific interests, such as nursery rhymes, cartoons / animations for kids, educational videos for kids, video games without inappropriate content, kids' toys demonstrations and ratings, children's music or dance performances, reality shows made for young children, and animal videos that are not inappropriate. A video is deemed appropriate if it has child-oriented activities and incentives, and simple language or content suitable for a wide audience.

**Inappropriate Content:** It includes content that is not suitable for young kids and preschoolers (under the age of 6) to consume. Violent, scary, or disturbing videos are some examples of inappropriate content. Inappropriate videos may include inappropriate visual content, language, or both.

**Irrelevant Content:** It includes content that is irrelevant or uninteresting to young viewers, such as building constructions, tax services, car purchases and services, politics, professional services, etc.

### 3.2 Young-kid-oriented Videos

In this section, we describe our collection of a large dataset of young-kid-oriented videos on YouTube.

**Collecting young-kid-oriented YouTube videos:** Common Sense Media [33] published a study of 1.6 K videos, reported by parents as watched by their children. But this video dataset is not available to the public. In another study, Papadamou et al. [24] released a dataset of 4.8 K videos that only includes 1.5 K young-kid-oriented videos. Moreover, 30% of these videos that were available when Papadamou et al. [24] performed their research have since been removed. Both datasets are too small to be representative of young-kid-oriented videos on YouTube. Hence, we create a large and representative dataset of young-kid-oriented YouTube videos. Since manual identification and labeling of videos is time-intensive, we decided to manually identify YouTube *channels* that may host a certain kind of appropriate content for young audiences. Then all videos within a given channel would be labeled with the channel-wide label. Our manual identification starts by using child-specific search terms (e.g., "nursery rhymes") and identifying candidate channels from search results.

**Manual Annotation of Videos:** Then, we randomly select ten videos from each channel and manually label them as appropriate, inappropriate or irrelevant. Our annotators manually review selected videos by inspecting the following data descriptors: (1) channel content, (2) video content, (3) channel titles, (4) video titles, (5) thumbnails, and (6) tags in YouTube. Five annotators label each selected video as appropriate, inappropriate or irrelevant. If all ten videos are labeled as appropriate, we keep this channel and add all the videos of the channel to our dataset D1. Our dataset, D1, contains 51 young-kid-oriented channels and 24.6 K videos.

**Inter-annotator Agreement:** We compute the agreement rate among these five annotators using Bennett et al.'s *S score* [16, 32]. The Bennett et al.'s *S score* value that we get is 0.96, which indicates a strong agreement across raters. We choose to use Bennett et al.'s *S score* as it is one of the widely used techniques for calculating inter-annotator agreement for more than two raters, as in our case. It accounts for the percentage of rater agreement that might be expected by chance, instead of just the simple agreement between raters, as with Cohen's Kappa coefficient [21].

**Downloading Videos:** Each YouTube video and channel has a unique identification code that can be used to download all information about it using YouTube Data API [5]. Thus, we extract all the video ids from the 51 channels in dataset D1 and automate video downloads. However, please note that YouTube Data API does not return any information about the ads shown on the YouTube videos.

### 3.3 Ads on YouTube Videos

In this section, we discuss the different types of ads that are shown on YouTube videos. And we download and annotate the ads for all videos in dataset D1. Further, we use our annotations to quantify the percentage of inappropriate ads shown alongside young-kid-oriented videos and present our findings in Section 4.

**Categories of Ads on YouTube:** According to YouTube Help [1], the major categories of ads on YouTube are (See Figure 1 for their examples):

**Video Ads or In-stream Ads:** is the most common category of YouTube ads that play before, during, or after the actual video is played [23, 35]. These are sponsored videos that appear in monetized organic content [27]. These may be skippable (a short video that viewers can skip after 5 seconds), non-skippable (a short video lasting up to 15 seconds and cannot be skipped), bumper ads (a short video lasting up to 6 seconds which viewers cannot skip) and in-feed ads (a short video that gets displayed on a user's homepage).

**Sidebar ads:** are images or video units, that are displayed outside, on the right of the actual video. Users can interact with these ads or turn them off at will. Clicking on one of these ads will take users to a landing page assigned by the advertiser.

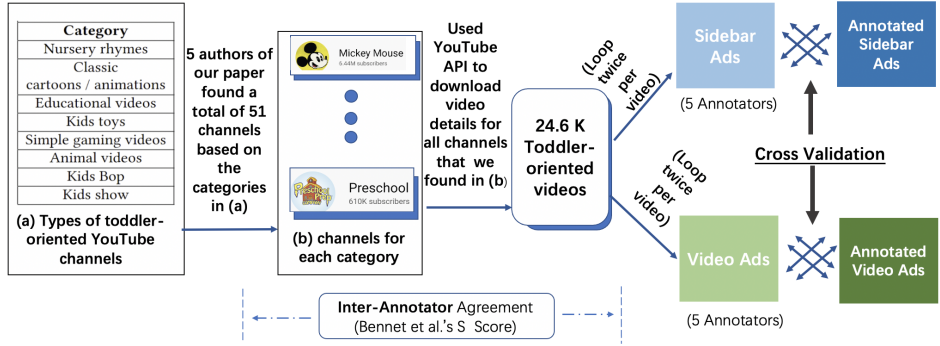

**Figure 4.** Architecture diagram of our methodology

**Masthead ads:** are ads displayed at the top of the home feed on YouTube. These are reserved mainly by one advertiser per country per day [28].

**Paid-promo ads:** are ads from a sponsor of a video, which are embedded within the content of the video.

**Overlay ads:** are small banner-like ads that take up about the bottom 20% of a video's screen without obstructing the user's view. These ads can contain images or text, and users can interact with them or turn them off at will. Clicking on one of these ads will take users to a landing page assigned by the advertiser.

In this paper, we focus on two of the most frequent categories of ads [23, 35] on any YouTube video page, i.e., the video ads and the sidebar ads. We leave the remaining categories of ads as future work.

### 3.4 Video Ads

This section describes how we collect video ads from dataset D1, annotate these ads, and validate our annotations.

**Collecting video ads from kid-oriented videos:** We use Selenium [4] scripts to play the appropriate videos from dataset D1, so we can scrape the video ads from the YouTube pages of appropriate videos. The dynamically rendered HTML content of a video Web page gives us the details about the video ads, if present, that are played on the actual video. Finally, we obtain 3,517 unique video ads after loading every video Web page from dataset D1 more than once. We refer to this list of video ads as the dataset, D2. Please note that YouTube may show different ads for a video on reloading the same YouTube video page again. Therefore, in this paper, we even consider that for a young-kid-oriented video, we cannot rule out the possibility of not seeing an inappropriate ad on reloading the video Web page if we did not see one in the past. Therefore, we load every video Web page more than once and capture the different ads.

**Annotating video ads:** Five authors of this paper each independently labeled roughly 703 out of 3,517 video ads from the dataset, D2, as appropriate, inappropriate, or irrelevant (see Section 3.1 for their definitions). Every annotator manually opened the assigned YouTube ad videos and watched the entire ad content to classify the video as appropriate, inappropriate, or irrelevant (see Figure 5).

**Annotation validation:** Since each ad is annotated just by a single annotator, it is necessary to verify the accuracy of the annotations. D2 contains a large number of ads, requiring significant manual effort for validation. Because we did not have more resources to do a large-scale annotation validation, hence, two other authors of the paper, who did not participate in our video ads annotation work, randomly selected 100 video ads each from the dataset D2. They independently validated the annotations for the randomly selected video ads. The aggregate accuracy of the annotation work, as reported in this annotation validation step, is 97%.

### 3.5 Sidebar Ads

This section describes how we collect sidebar ads from dataset D1, annotate these ads, and validate our annotations.

**Collecting sidebar ads from kid-oriented videos:** Analogous to the ad collection step for video ads (Section 3.4), we use the dynamically rendered HTML content to fetch sidebar ads details, including the ad (1) title, (2) description, and (3) Website URL. Finally, we obtain 1,069 sidebar ads after playing every video from dataset D1 more than once. We refer to this list of sidebar ads as the dataset, D3.

**Annotating sidebar ads:** Five authors of the paper each independently labeled roughly 213 out of 1,069 sidebar ads from the dataset D3 as appropriate, inappropriate, or irrelevant. Additionally, every annotator manually checked the details of each sidebar ad, including the referenced website for any given ad. They independently assigned appropriate, inappropriate, or irrelevant labels for the sidebar ads: (1) as visible on the YouTube page, and (2) embedded URLs in the ad (see Figures 2 and 3). We assign separate labels for sidebar ad visibility on YouTube pages and embedded URLs in the sidebar ads because sidebar ads can be deceiving too. Clicking on sidebar ads may open Web pages that contain different content than advertised. In this work, we refer to a sidebar ad that looks appropriate from its description and title, but is inappropriate from the actual content in its linked URL as **deceptive sidebar ads** (see Figure 6).

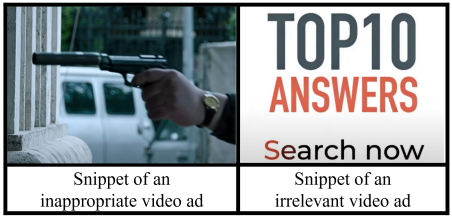

**Figure 5.** Snippet of an inappropriate video ad and an irrelevant video ad

**Annotation validation:** Analogous to the annotation validation step for video ads (Section 3.4), two other authors of the paper, who did not participate in the ads annotation work, randomly selected 100 sidebar ads each from the dataset D3. They independently validated the annotations of the randomly selected sidebar ads for their descriptions and their referenced websites. And the aggregate accuracy of the annotation work, as reported in this annotation validation step, is 97.5%.

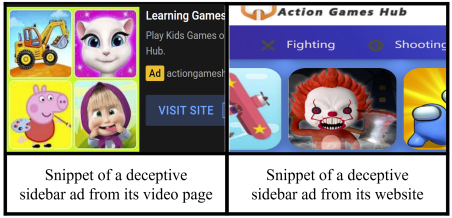

**Figure 6.** Snippet of a deceptive sidebar ad

### 3.6 Ethics

We only collect publicly available data on the Web and do not (1) interact with online users, nor (2) imitate any logged-in activity on YouTube or other platforms. Therefore, the IRB approval was not required for this work.

## 4 Findings

This section discusses our findings of YouTube ads displayed alongside young-kid-appropriate videos, using the datasets D1, D2, and D3 (shown in Table 1). We find that 9.9% of the 4.6 K unique advertisements shown on these 24.6 K videos contain inappropriate content for young children (shown in Table 4). Moreover, we observe that 26.9% of all the 24.6 K appropriate videos included at least one ad that is inappropriate for young children (shown in Table 5). Our findings complement the observations of Common Sense Media [9] about inappropriate ads displayed on young-kid-oriented videos. These organizations found that 20% of the ads that young kids watch on YouTube are inappropriate, on a sample of 1.6 K videos. Comparatively, our findings are based on a more recent and much larger dataset, which accounts for some differences in our findings.

**Table 1.** Details about our datasets

| Dataset D1 | Dataset D2 | Dataset D3 |
|---|---|---|
| 24,592 appropriate videos | 3,517 video ads | 1,069 sidebar ads |

### 4.1 Video Ads

We find that 4.5% (160 out of 3,517) video ads that display on young-kid-oriented videos are inappropriate for the young-kids to watch (shown in Table 2). On the other hand, only 7.4% (262 out of 3,517) ads are safe for kids to watch. Further, 88.0% (3,095 out of 3,517) ads are irrelevant for young children. Also, gaming involving physical violence is the most common category of inappropriate video ads shown on young-kid-oriented videos, and Online Software/Web Services are the most common category of irrelevant video ads.

**Table 2.** Video ads from 24.6 K young-kid-oriented videos

| Unique | Appropriate | Inappropriate | Irrelevant | Total |
|---|---|---|---|---|
| Ad | 262 | 160 | 3,095 | 3,517 |
|  | 7.4% | 4.5% | 88.0% |  |

**Table 3.** Sidebar ads from 24.6 K young-kid-oriented Videos

| Unique | Appropriate | Inappropriate | Irrelevant | Total |
|---|---|---|---|---|
| Ad Description | 517 | 3 | 549 | 1,069 |
|  | 48.3% | 0.3% | 51.4% |  |
| Ad URL | 217 | 292 | 560 | 1,069 |
|  | 20.3% | 27.3% | 52.4% |  |
| Ad Description/URL | 217 | 294 | 558 | 1,069 |
|  | 20.3% | 27.5% | 52.2% |  |

**Table 4.** Total ads from 24.6 K young-kid-oriented Videos

| Unique | Appropriate | Inappropriate | Irrelevant | Total |
|---|---|---|---|---|
| Video ads + Sidebar ads | 479 | 454 | 3,653 | 4,586 |
|  | 10.4% | 9.9% | 79.7% |  |

**Table 5.** Ads shown on 24.6 K appropriate videos that influence those videos

| Unique | Appropriate | Inappropriate | Irrelevant | Total |
|---|---|---|---|---|
| Video ads | 7,858 | 1,356 | 15,378 | 24,592 |
|  | 32.0% | 5.5% | 62.5% |  |
| Sidebar ads | 15,548 | 5,578 | 3,466 | 24,592 |
|  | 63.2% | 22.7% | 14.1% |  |
| Video ads + Sidebar ads | 4,944 | 6,610 | 13,038 | 24,592 |
|  | 20.1% | 26.9% | 53.0% |  |

## 4.2 Sidebar Ads

We find that 27.5% (294 out of 1,069) of all the sidebar ads that are shown on young-kid-oriented videos are inappropriate for the young-kids to watch (shown in Table 3). 20.3% (217 out of 1,069) ads are suitable for young children to watch. Further, 52.2% (558 out of 1,069) ads are irrelevant. Gaming involving physical violence, again, is the most common category of inappropriate sidebar ads, and Online Software/Web Services is also the most common category of irrelevant sidebar ads.

Just based on the description and title of the sidebar ads, 48.3% (517 out of 1,069) are appropriate, 0.3% (3 out of 1,069) are inappropriate, and 51.4% (549 out of 1,069) are irrelevant. However, based on the referenced Web pages of sidebar ads, 20.3% (217 out of 1,069) are appropriate, 27.3% (292 out of 1,069) are inappropriate, and 52.4% (560 out of 1,069) are irrelevant. Thus we find that 26% (284 out of 1,069) of the sidebar ads are deceptive (see Section 3.5).

## 4.3 Findings about inconsistencies in ad reporting

As per YouTube Help [2], if one finds an ad inappropriate or violating Google's ad policies, one can report the ad to YouTube directly using the "Why this ad?" button. On clicking this button, one can select the "Report this ad" (to complain about the ad to YouTube that the ad is not appropriate for all [15]), "Stop seeing this ad" option (for users who do not wish to see the ad [14]), or both. We find inconsistencies in the ad reporting feature (see Figure 7). We visited 200 different YouTube video pages and found that the reporting options were either completely disabled or incomplete (i.e., only the "Report this ad" option was available but not the other) for half of our total trials.

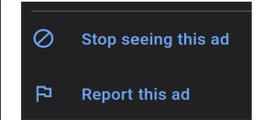

**Figure 7.** Inconsistencies in YouTube video ad reporting

## 5 Recommendations

In this section, we discuss what YouTube can do to avoid displaying inappropriate or irrelevant ads to young children.

**Serve different ads, alongside the video, that target different age groups:** We recommend YouTube to serve distinct ads for different demographics. For example, if one is watching a young-kid-oriented video, YouTube should serve ads that only target young kids.

**Identifying content inappropriate for certain audience groups:** While our present datasets focus on identifying ads that are inappropriate for young children, comparable datasets might be produced for any audience.

**Monitor ads that link to external Webpages:** Based on our findings of deceptive sidebar ads in Section 4.2, we recommend that YouTube should monitor the embedded content of sidebar ads or any other ads that link to external Web pages.

**Introducing restricted mode on YouTube ads:** Similar to the restricted mode for YouTube videos, YouTube should publish a restricted mode for ads that can remove inappropriate content for certain vulnerable groups.

**Ad reporting:** Based on our findings about ad reporting inconsistencies in Section 4.3, we recommend YouTube to enable full ad reporting functionality for all ads.

## 6 Future Work

Our future work is to automate the annotation work. Video ads automatic classification can be achieved using existing techniques such as the ones proposed by Papadamou et al. [24] and Tahir et al. [30], to determine whether the video ad is appropriate or inappropriate for kids. For sidebar ads, we need even larger and balanced datasets comprising enough inappropriate and appropriate ad samples for training.

## 7 Conclusion

YouTube has gained popularity among young children too, over the years. Unfortunately, the prevalence of inappropriate ads in young-kid-oriented videos has become a serious issue. In this paper, we conducted a large-scale analysis of 24.6 K YouTube videos that are appropriate for young kids. We found that 9.9% of the 4.6 K unique advertisements shown on these 24.6 K videos included inappropriate content for young audiences. Further, we discovered that 26.9% of all the 24.6 K appropriate videos had at least one inappropriate ad for young children. Finally, we presented multiple recommendations for future directions to solve this issue.